\documentclass[aps,prl,reprint,groupedaddress,showkeys,superscriptaddress]{revtex4-1}
\usepackage{graphicx}
\usepackage{dcolumn}
\usepackage{bm}
\usepackage{epstopdf}
\usepackage{amsmath}
   \usepackage[all]{xy}
\usepackage{subfigure}
\usepackage{color}
\begin{document}


 \title{Inferring energy dissipation from violation of  the Fluctuation-Dissipation Theorem}

 \author{Shou-Wen Wang}
  \email{wangsw09@csrc.ac.cn}
\affiliation{Beijing Computational Science Research Center, Beijing, 100094, China}
\affiliation{Department of Engineering Physics, Tsinghua University, Beijing, 100086, China}

\date{\today}
  \graphicspath{{./figure/}}
\begin{abstract}

The Harada-Sasa equality  elegantly  connects the energy dissipation rate of a moving object with its measurable violation of the Fluctuation-Dissipation Theorem (FDT).   Although proven  for Langevin processes,  its validity remains unclear for \emph{discrete} Markov systems whose forward and backward transition rates respond \emph{asymmetrically} to  external perturbation.   A typical example is a motor protein called kinesin.    Here we show generally that the FDT violation persists surprisingly in the high-frequency limit due to the  asymmetry, resulting in a divergent FDT violation integral and thus a complete breakdown of the Harada-Sasa equality.   A renormalized FDT violation integral still well predicts the dissipation rate when each discrete transition produces a small entropy in the environment.   Our study also suggests a new way to infer this perturbation asymmetry based on the measurable high-frequency-limit FDT violation. 


\end{abstract}

\maketitle

\emph{Introduction.--}Recent development of technology has allowed direct observation and control of molecular fluctuations,  thus opening up a new field to explore  nano machines that operate out of equilibrium~\cite{toyabe2015nonequilibrium,martinez2017colloidal,ciliberto2017experiments}.    An important approach to investigate a stochastic system is to study both its spontaneous fluctuation and the elicited response to perturbation.   For the recorded velocity $\dot{x}_t$ of a particle (with $x_t$ being its position at time $t$),   its spontaneous fluctuation is  captured by the temporal correlation function:  $C_{\dot{x}}(t-\tau)\equiv \langle (\dot{x}_t-\langle \dot{x}\rangle_{ss})(\dot{x}_\tau-\langle \dot{x}\rangle_{ss})\rangle_{ss}$,  with $\langle \cdot\rangle_{ss}$ denoting the average over the stationary ensemble.   On the other hand,   the velocity response  to a small external force $h$  is captured by the temporal response function determined from the functional derivative $R_{\dot{x}}(t-\tau)\equiv \delta \langle \dot{x}_t\rangle/\delta h_\tau$.      For equilibrium systems,  these two functions are closely related through the fundamental Fluctuation-Dissipation Theorem (FDT)~\cite{kubo1966fluctuation},  which in the Fourier space reads
 \begin{equation}
 \tilde{C}_{\dot{x}}(\omega)=2Tk_B\tilde{R}_{\dot{x}}'(\omega),
 \end{equation}
 where prime denotes the real part, $T$ is the bath temperature, and the Boltzmann factor $k_B$ is set to be 1 hereafter.   Violation of the FDT has been widely used to characterize non-equilibrium systems, including glassy systems~\cite{cugliandolo1997fluctuation,LeticiaPREtemperature},  hair bundles~\cite{martin2001comparison}, and  cytoskeleton networks~\cite{Mizuno2007cytoskeletonNetwork}.  
 
 The generalization of the FDT for systems in non-equilibrium steady state  has been studied intensively~\cite{baiesi2009fluctuations,seifert2010fluctuation,baiesi2013update,speck2006restoring,prost2009generalized}.   
In particular, for systems described by Langevin equations,  Harada and Sasa have shown that the violation integral of the FDT gives the dissipation rate $\dot{q}$ for the observed variable $x$~\cite{harada2005equality,harada2006energy,harada2007fluctuations}:
   \begin{equation}
I\equiv \langle\dot{x}\rangle_{ss}^2+ \int_{-\infty}^\infty \Big[ \tilde{C}_{\dot{x}}(\omega)-2T\tilde{R}_{\dot{x}}'(\omega)\Big]\frac{d\omega}{2\pi}=\frac{\dot{q}}{\gamma},
\label{eq:HS1}
 \end{equation}
with $\gamma$  the friction coefficient.   The Harada-Sasa (HS)  equality  has been applied successfully to infer the energetics of F1-ATPase, a rotary motor protein~\cite{toyabe2015single,toyabe2010nonequilibrium}.  Our recent study demonstrated that it is also useful for inferring hidden dissipation of  timescale-separated systems when having access to only  slow variables~\cite{Wang2016entropy,Wang2016FRRviolation}.  Eq.~(\ref{eq:HS1}) has also been generalized to more elaborated Langevin systems~\cite{DeutschNarayan2006,fodor2016far,nardini2017entropy}.

 Although the  HS equality seems very general,  its validity  remains unclear  for \emph{discrete} Markov processes.   In this context,   Lippiello \emph{et al.} have shown  that  the HS equality is recovered  when  entropy production in the environment  is small for each jump~\cite{Lippiello2014fluctuation}.  A central assumption there is  that  the forward and backward transition rates respond \emph{symmetrically} to the external perturbation.  However,  this symmetry is violated for molecular motors, according to recent experimental and modeling work~\cite{zimmermann2012efficiencies,kawaguchi2014nonequilibrium,Zimmermann2015motors,brown2017allocating,taniguchi2005entropy,ariga2017nonequilibrium}.   Furthermore,  various forms of generalized FDT that go beyond symmetric perturbation  reveal non-trivial dependence on the asymmetry~\cite{diezemann2005fluctuation,maes2010response,seifert2010fluctuation}, in sharp contrast with the simplicity of the HS equality.  
 
  Here, we clarify  the connection between dissipation rate and violation of the FDT for Markov systems with perturbation asymmetry.  We find surprisingly that the FDT is violated even in the high-frequency limit, leading to a divergent FDT violation integral,  although the dissipation rate remains finite.  We propose two renormalization schemes to remove the divergence of the FDT violation integral, and show that the renormalized integrals well predict the dissipation rate when the entropic change per jump is small.   The main results are illustrated with a minimum model for kinesin.

 \emph{General Markov systems.--}Consider a general Markov process with $N$ states.  The transition from state $n$ to state $m$ ($1\le n,m\le N$) happens with  rate $w_n^m$.  The probability  $P_n(t)$ at state $n$ and time $t$ evolves according to the following master equation
\begin{equation}
 \frac{d}{dt}P_n(t)=\sum_m M_{nm}P_m(t),
 \label{eq:govMatrix}
 \end{equation}
 where $M$ is assumed to be  an irreducible transition rate matrix determined by $M_{nm}=w_m^n-\delta_{nm}\sum_k w_n^k$, with $\delta_{nm}$ being the Kronecker delta.      The $j$-th left and right eigenmodes,  denoted as $x_j(n)$ and $y_j(n)$ respectively,  satisfy the characteristic equations  $\sum_m M_{nm}x_j(m)=-\lambda_j x_j(n)$ and  $\sum_m y_j(m)M_{mn}=-\lambda_j y_j(n)$. Here, the minus sign is introduced to have an ``eigenvalue" $\lambda_j$ with a positive real component~\cite{van1992stochastic}.   These eigenvalues are arranged in the ascending order by their real part, i.e., $\text{Re}(\lambda_1)\le \text{Re}(\lambda_2)\le \cdots$.     This system has a unique stationary distribution $P_{m}^{ss}$ that satisfies $\sum_m P_m^{ss}=1$.  For the ground state associated with    $\lambda_1=0$,    $y_1(n)$ should be constant and $x_1(m)$ be proportional to $P_m^{ss}$.  Here,  we fix  $y_1=1$ and  $x_1(m)=P_{m}^{ss}$.  For this system,  we can always find a  set of   eigenmodes that satisfy the  orthogonal relations $\sum_m x_j(m)y_{j'}(m)=\delta_{jj'}$ and completeness relations $\sum_j x_j(n)y_j(m)=\delta_{nm}$,  which we use in the following analysis.     The left and right eigenmodes are coupled  for equilibrium systems:  $x_j(m)=y_j(m)P_{m}^{eq}$.  This is not true   for non-equilibrium systems.

We introduce an external perturbation  $h$ that   modifies the transition rates to be 
\begin{equation}
 \tilde{w}_m^n=w_m^n\exp\left[h(\theta_n^m+\frac{1}{2}) \frac{\mathcal{Q}_n-\mathcal{Q}_m}{T}\right].
\label{eq:specific}
\end{equation}
Here,  $\mathcal{Q}_m$ is a  variable  conjugate to perturbation $h$,  and $\theta_n^m$ parameterizes the asymmetry of the transition rates in response to external perturbation.  $\theta_n^m$ may vary for different transitions, but should satisfy $\theta_n^m=-\theta_m^n$.   $\theta_n^m=0$  corresponds to the symmetric case.   We are interested in the correlation and response spectrum of the velocity observable   $\dot{Q}_t=d\mathcal{Q}_{n_t}/dt$.   The strategy is to project these spectra onto the eigenspace.  We introduce the projection coefficients:   $\alpha_j\equiv\sum_n \mathcal{Q}_nx_j(n)$, $\beta_j \equiv \sum_{n}\mathcal{Q}_{n}  y_j(n)P_n^{ss}$,  and $\phi_j \equiv \sum_{n} B_{n} y_j(n)$,  where $B_n$  captures the effect from perturbation, and is given by  $B_n= \sum_{m} ( \theta_n^mJ_n^m+\mathcal{A}_n^m)(\mathcal{Q}_n-\mathcal{Q}_m)/T$.
Here,  $\mathcal{A}_n^{m}\equiv  (w_n^m P_n^{ss}+w_{m}^n P_{m}^{ss})/2$  is the dynamical activity between state $n$ and $m$,   while $J_n^{m}\equiv  w_n^m P_n^{ss}-w_{m}^n P_{m}^{ss}$ is the net flux from state $n$ to $m$. 
 Then,  we obtain
 \begin{subequations}\label{eq:FDT-expansion}
 \begin{eqnarray}
  \tilde{C}_{\dot{Q}}(\omega)&=&\sum_{j=2}^N 2\alpha_j\beta_j\lambda_j \Big[1-\frac{1}{1+(\omega/\lambda_j)^2}\Big], \label{eq:Cvelo_fre}\\
     \tilde{R}_{\dot{Q}}(\omega)&=&\sum_{j=2}^N \alpha_j\phi_j\Big[1- \frac{1+ i(\omega/\lambda_j) }{1+(\omega/\lambda_j)^2}\Big], \label{eq:Rvelo_fre}
 \end{eqnarray}
 \end{subequations}
 with $i$ the imaginary unit.    We have used  this framework previous in the context of symmetric perturbation~\cite{Wang2016entropy,Wang2016FRRviolation}.    See Supplemental Material~\cite{supp} for more details.

Let us consider the high frequency limit  first.  According to Eq.~(\ref{eq:FDT-expansion}),  we have $\tilde{C}_{\dot{Q}}(\infty) =\sum_{j=2}^N 2\alpha_j\beta_j\lambda_j$ and $ \tilde{R}_{\dot{Q}}(\infty)=\sum_{j=2}^N \alpha_j\phi_j$.  Following the definitions of these coefficients,  we obtain
 \begin{subequations}\label{eq:high-FDT-general}
\begin{eqnarray}
\tilde{C}_{\dot{Q}}(\infty) &=&  \sum_{n,m} (\mathcal{Q}_n-\mathcal{Q}_m)^2 \mathcal{A}_n^m,\\
\tilde{R}_{\dot{Q}}(\infty) &=& \frac{1}{2T} \sum_{n,m} (\mathcal{Q}_n-\mathcal{Q}_m)^2 \Big(\theta_n^m J_n^m+ \mathcal{A}_n^m\Big).\quad
\end{eqnarray}
\end{subequations}
In obtaining Eq.~(\ref{eq:high-FDT-general}a), we note that $\sum_j x_j(n)\lambda_j y_j(m)=-M_{nm}$,  and that any summation over the full state space is invariant under the  switching of the label, i.e.,  $n\leftrightarrow m$.  Because $\theta$ is introduced only at the stage of perturbation here,   the correlation spectrum does not depend on $\theta$.     More specifically,  $\tilde{C}_{\dot{Q}}(\infty) $   only depends on the  activity $\mathcal{A}_n^m$,  while  $\tilde{R}_{\dot{Q}}(\infty) $ has an additional dependence on the flux $J_n^m$ in the presence of an asymmetric load-sharing factor. 
The  FDT violation in the high-frequency limit  is then
  \begin{equation}   \label{eq:lambda-beta-phi-2}
 \mathcal{V}_{\infty}\equiv \lim_{\omega\to \infty} \Big[ \tilde{C}_{\dot{x}}(\omega)-2T\tilde{R}_{\dot{x}}'(\omega)\Big]=-\sum_{n,m} \theta_m^n J_m^n (\mathcal{Q}_m-\mathcal{Q}_n)^2.
  \end{equation} 
 It vanishes for any equilibrium systems ($J_m^n=0$) or non-equilibrium systems with symmetric perturbation ($\theta_m^n=0$).   Otherwise, a finite FDT violation persists even in the high-frequency limit, which is quite surprising.   When the transitions are dominated by futile back-and-forth jumps, i.e., $|\mathcal{A}_n^m|\gg |J_{n}^m|$,  the system has a relatively small high-frequency-limit violation, i.e., $|\mathcal{V}_{\infty}/\tilde{C}_{\dot{Q}}(\infty)| \ll 1$.  This will be the typical case when individual jumps produce a small entropic change in the environment, as will be illustrated later.

The direct consequence of a non-zero $\mathcal{V}_\infty$ is a divergent FDT violation integral $I$ and thus complete breakdown of the HS equality (as the dissipation rate still remains finite).   To get rid of divergence,  we first subtract  $\mathcal{V}_{\infty}$ from the violation spectrum and then introduce the renormalized FDT violation integral: 
\begin{equation}
I_*\equiv  \langle \dot{Q}\rangle_{ss} +\int_{-\infty}^\infty \Big[\tilde{C}_{\dot{Q}}(\omega)-2T\tilde{R}'_{\dot{Q}}(\omega)-\mathcal{V}_\infty\Big]\frac{d\omega}{2\pi}.
\end{equation} 
A more practical scheme of renormalization will be discussed towards the end.   Combined with Eq.~(\ref{eq:FDT-expansion}),   we  obtain $ I_*=\sum_j \lambda_j\alpha_j \left(T\phi_j-\beta_j\lambda_j\right)$~\cite{note}.  Using the definitions of these coefficients and summing over all eigenmodes,  we obtain~\cite{supp}
   \begin{equation}
I_* =\sum_{n,m} \Big( \frac{\bar{\nu}_n+\bar{\nu}_m}{4}+\frac{\bar{\nu}_n-\bar{\nu}_m}{2}\theta_n^m\Big)J_n^m (\mathcal{Q}_m-\mathcal{Q}_n),
\label{eq:lambda-beta-phi-3}
\end{equation}
where  $\bar{\nu}_n\equiv \sum_m w_n^m (\mathcal{Q}_m-\mathcal{Q}_n)$ is the average change rate of $Q_t$  when it starts from state $n$.   Evidently from this equation, the  FDT violation only comes from transitions that change the observable $\mathcal{Q}_n$,  as it should,  and it is proportional to the local net flux $J_n^m$,  the signature of non-equilibrium systems.   Below, we discuss the structure of $\mathcal{V}_\infty$ and the connection between $I_*$ and the dissipation rate through more specific models.

%

 \begin{figure}
 \centering
 \includegraphics[width=8.5cm]{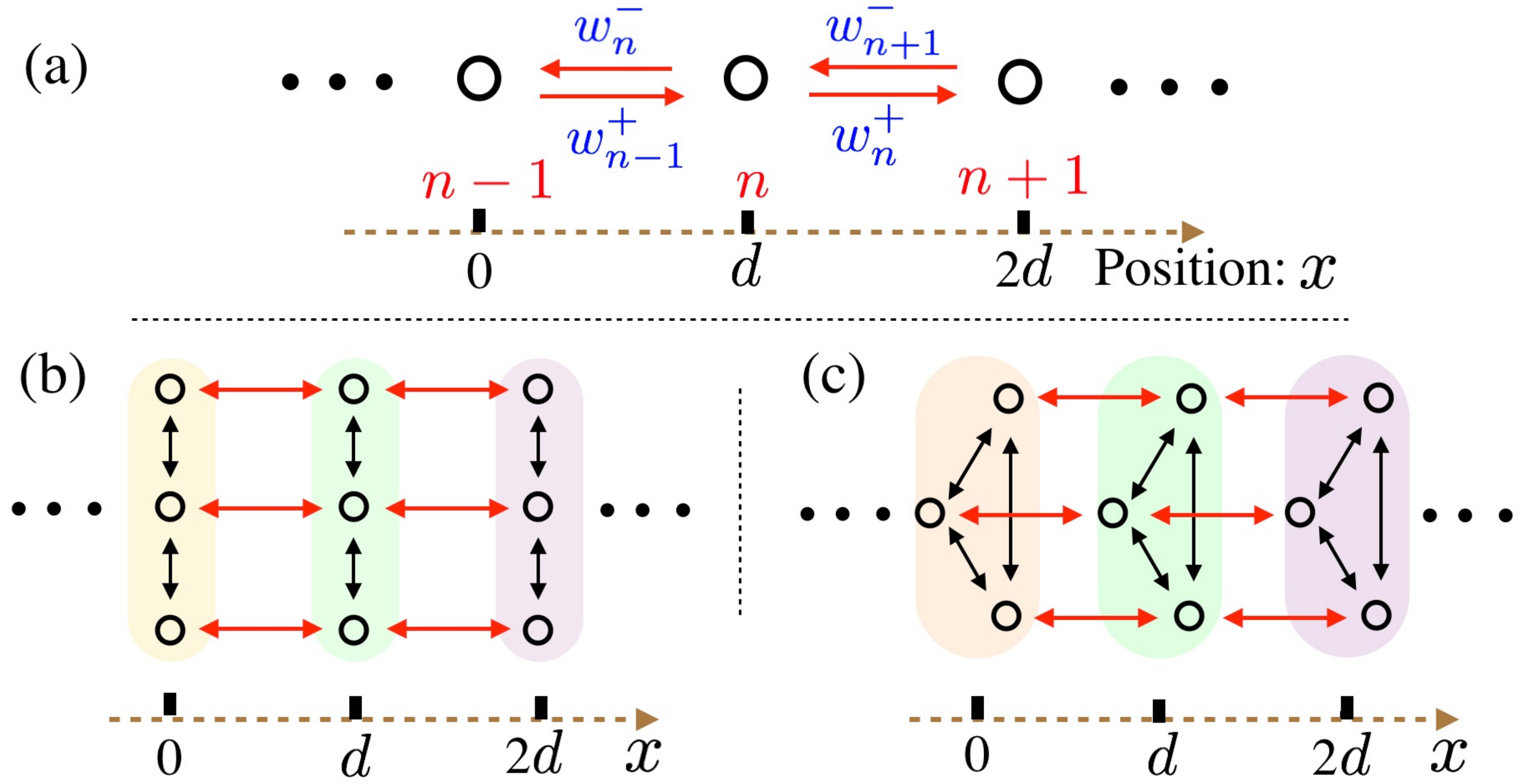}
 \caption{ (a) One-dimensional (1-d) hopping process.  (b)(c)  Multi-dimensional hopping models.  The corresponding observable $\mathcal{Q}_n$, which is $x$ here, does not distinguish microscopic states within each colored block.  As with the 1-d hopping model,  we assume that $\theta$ is the same for all red transitions.  These models may describe molecular motors that hop along a discrete lattice with several internal chemical states.  They also resemble sensory adaptation model in E.\emph{coli}~\cite{shouwen2015adaptation}.   }
 \label{fig:continuum}
 \end{figure}

\emph{Application to various models.--}
Consider  a particle hopping along a discrete lattice with a lattice constant $d$, as illustrated in FIG.~\ref{fig:continuum}(a).      Each state $n$ has a well-defined energy $U_n$.   The transition rates are assumed to satisfy 
 \begin{subequations}\label{eq:w-theta-1d}
 \begin{eqnarray}
 w_n^+&=&w_0\exp\Big((\theta+\frac{1}{2})\frac{ \Delta U_{n}+hd}{T} \Big), \\
 w_{n+1}^-&=&w_0 \exp\Big((\theta-\frac{1}{2}) \frac{\Delta U_{n}+hd}{T} \Big),
 \end{eqnarray}
 \end{subequations}
with $w_0$ the constant prefactor, and $\Delta U_{n}\equiv U_n-U_{n+1}$ the dissipation per jump.   This model satisfies local detailed balance, i.e.,  $w_n^+/w_{n+1}^-=\exp([ \Delta U_{n}+hd]/T)$.  We assume that  $U_n$ is constructed from a continuous function $U(x)$ via $U_n=U(nd)$.  The energy landscape can be tilted to drive the system out of equilibrium. 

Firstly, we derive the high-frequency violation $\mathcal{V}_{\infty}$.   For this system, the conjugate observable $\mathcal{Q}_n$ is position $x$.   We note that $(\mathcal{Q}_m-\mathcal{Q}_n)^2=d^2$ for all allowed transitions. Furthermore, both the flux $J=\langle \dot{x}\rangle_{ss}/L$ and the asymmetric factor $\theta$ are constant in the state space.  Therefore,  Eq.~(\ref{eq:lambda-beta-phi-2}) is reduced to  
 \begin{equation}
 \mathcal{V}_{\infty}=-2\theta \langle \dot{x}\rangle_{ss} d.
 \label{eq:asymmetry-vio}
 \end{equation}
   This simple relation~(\ref{eq:asymmetry-vio})   can be easily generalized to multi-dimensional hopping processes illustrated in  Fig.~\ref{fig:continuum}(b)(c),   by lumping states within each colored block and fluxes between two connected blocks.  For such multi-dimensional models,   $\mathcal{V}_{\infty}$ may vanish   even if the system remains out of equilibrium,   as  $\langle \dot{x}\rangle_{ss}=0$ is not a sufficient condition for equilibrium here.   This is not possible for 1-d systems.

Secondly, we derive the renormalized HS equality. 
According to  $I_*$ in Eq.~(\ref{eq:lambda-beta-phi-3}),  we have 
 \begin{equation}
I_*=d  \sum_n \Big( \frac{\bar{\nu}_n+\bar{\nu}_{n+1}}{2} +  \theta  (\bar{\nu}_n-\bar{\nu}_{n+1}) \Big)J_n^{n+1} .
\label{eq:I_*-1d}
 \end{equation} 
Here, $\bar{\nu}_n=d (w_n^+-w_n^-)$.    We introduce $\epsilon_n\equiv \ln [w_n^+/w_{n+1}^-]$ as the entropy produced in the environment per jump, which has a characteristic amplitude   $\epsilon$.     Expanding Eq.~(\ref{eq:I_*-1d}) in Taylor series of $\epsilon$,  we have $\bar{\nu}_n+\bar{\nu}_{n+1}= 2w_0d \epsilon_n+\theta O(\epsilon^2)+O(\epsilon^3)$ and $\bar{\nu}_n-\bar{\nu}_{n+1}=O(\epsilon^2)$~\cite{supp}, which gives
 \begin{equation}
  I_*= w_0d^2 \sum_n J_n^{n+1}\epsilon_n \Big(1+ \theta O (\epsilon )+O(\epsilon^2 ) \Big).
  \label{eq:I*-2}
  \end{equation}
We identify $\dot{q}=T \sum_n J_n^{n+1} \epsilon_n$ as  the dissipation rate of the stochastic trajectory $Q_t$, and $\gamma_*=T/(w_0 d^2)$ as the effective friction coefficient.    Finally,   we obtain  
  \begin{equation}
  \gamma_* I_*=\dot{q} \Big(1+ \theta O(\epsilon )+O (\epsilon^2 )\Big).
  \label{eq:GHS}
  \end{equation}
  The renormalized HS equality, i.e., $\gamma_* I_*=\dot{q}$, is recovered when $\epsilon$ is small, regardless of asymmetry and discreteness.  While the asymmetry leads to a first order deviation, the deviation of   discreteness  is only of the second order, thus much smaller.  The assumption of a constant $\theta$ is  crucial here.   Throughout the derivation, we did not assume that $J_n^{n+1}$ is constant,  a characteristic property of 1-d systems.  Hence, Eq.~(\ref{eq:GHS}) can be generalized  to multi-dimensional models  in  Fig.~\ref{fig:continuum}(b)(c), as discussed in Supplemental Material~\cite{supp}.

\begin{figure}
 \centering
 \includegraphics[width=8.5cm]{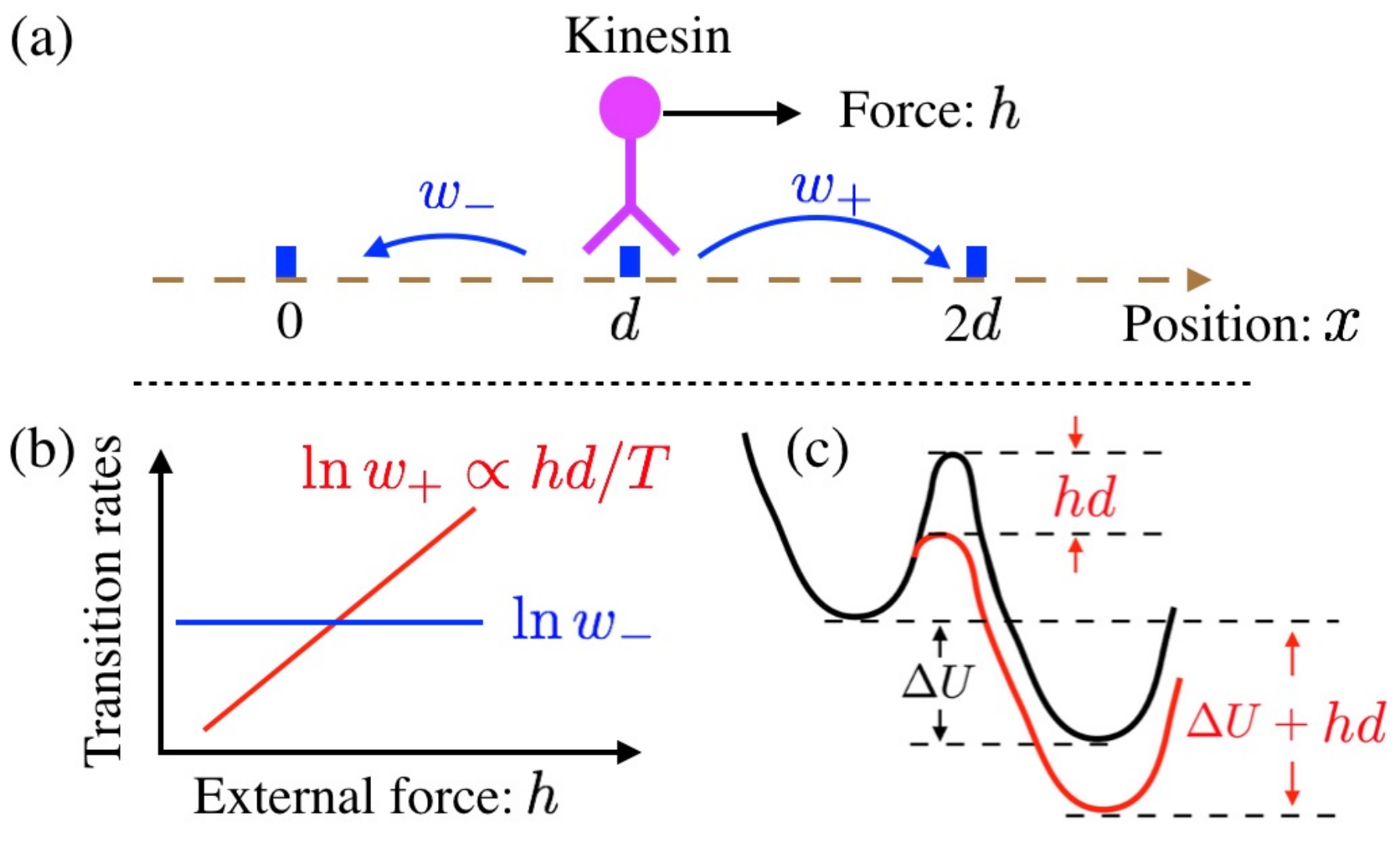}
 \caption{ (a) A simplified Markov model for kinesin.  (b) The experimentally suggested relations between transition rates  $w_\pm$ and the external force $h$~\cite{taniguchi2005entropy}, corresponding to  the case with $\theta=0.5$.  (c) The  energy landscape connecting neighboring states.   The external force only varies the energy barrier for the forward transition.  }
 \label{fig:motor1}
 \end{figure}

\emph{Minimum model for kinesin.--}A kinesin is a type of molecular motor that, powered by ATP, moves along microtubule filaments.  Following the experimental and modeling work in~\cite{taniguchi2005entropy},  we use the biased diffusive model  presented in Fig.~\ref{fig:motor1}(a) to describe the stepwise dynamics of this motor, with $d$ the step size.  This is a special case of the 1-d hopping model that has translational invariance.  The dissipation per jump $\Delta U$ can be tuned by changing ATP concentration, and  $h$ is the external force that is applied to the bead attached to the motor in a typical experimental setup.   Experiments show that the external force only affects the forward transition rate $w_+$~\cite{taniguchi2005entropy}, as illustrated in Fig.~\ref{fig:motor1}(b). This situation occurs when the external force only varies  the energy barrier for the forward transition  (Fig.~\ref{fig:motor1}c).   The scenario of kinesin corresponds to a completely asymmetric model with $\theta=0.5$.

 \begin{figure}
 \centering
 \includegraphics[width=8.5cm]{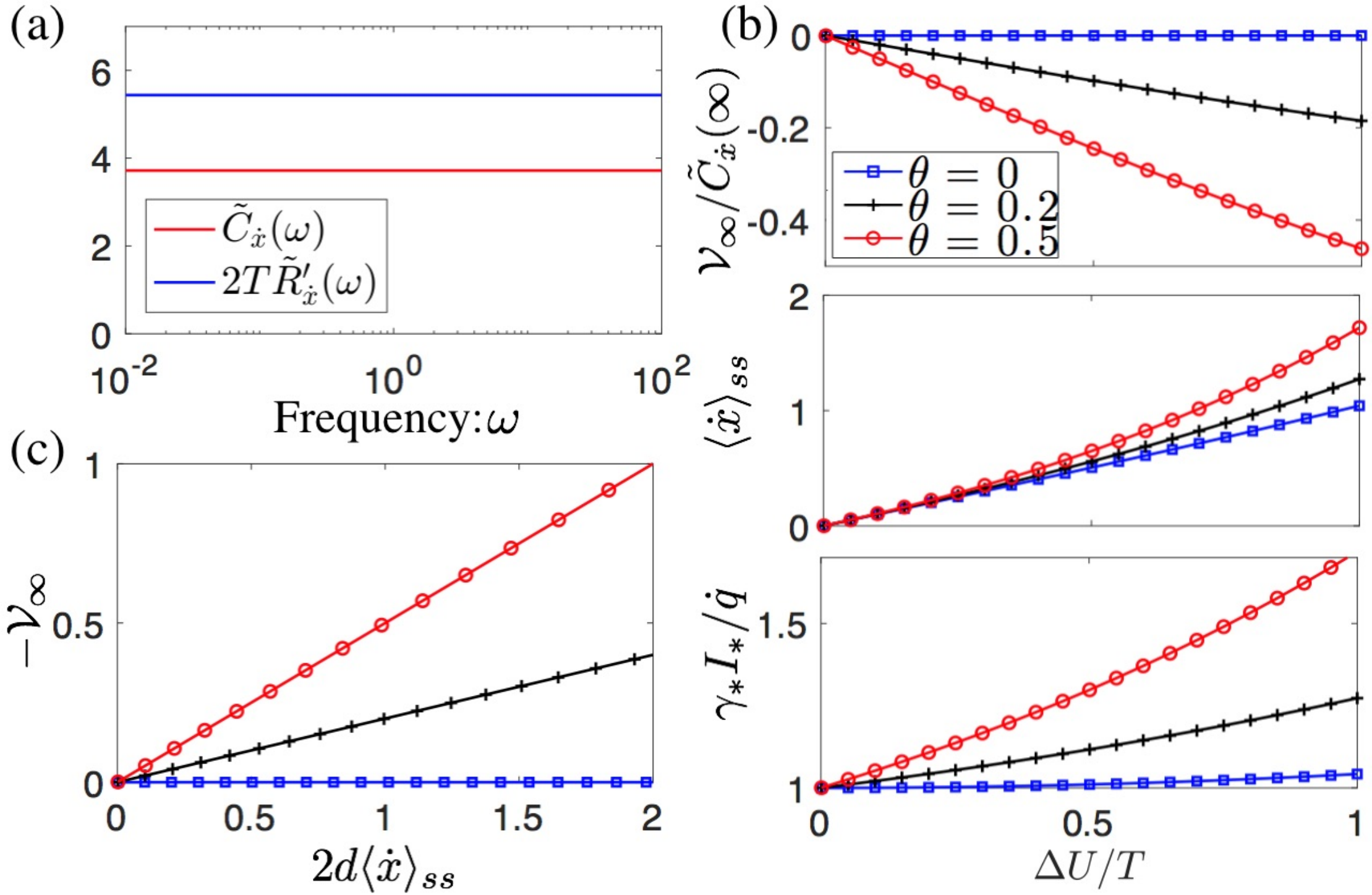}
 \caption{  (a) The correlation spectrum $\tilde{C}_{\dot{x}}(\omega)$ and the (real part of)  response spectrum $\tilde{R}_{\dot{x}}'(\omega)$ for the velocity $\dot{x}_{t}$, obtained at $\Delta U=1$ and $\theta=0.5$.  A finite violation of FDT, $\mathcal{V}_\infty$, persists even in the high frequency limit.  (b) 
  The relative  violation of the FDT in the high-frequency limit [$\mathcal{V}_{\infty}/\tilde{C}_{\dot{x}}(\infty)$],  the average drifting velocity $\langle \dot{x}\rangle_{ss}$, and the predicted dissipation rate $\gamma_* I_*$ based on the renormalized FDT violation integral  against the actual rate $\dot{q}$.  Here, the control parameter is the entropic change per jump, i.e., $\Delta U/T$.   (c) Verification of Eq.~(\ref{eq:asymmetry-vio}).  The slope of each curve gives the corresponding $\theta$.    Other parameters:   $w_0=1$,  $d=1$, and $h=0$.   }
 \label{fig:motor2}
 \end{figure}

This simple model allows analytical solutions.  At the steady state with $h=0$,  the average velocity is given by $\langle \dot{x}\rangle_{ss}= d(w_+-w_-)$, and the dissipation rate  $\dot{q}=\Delta U\langle \dot{x}\rangle_{ss}/d=\Delta U(w_+-w_-)$. The correlation spectrum of the velocity $\dot{x}$ is found to be 
$\tilde{C}_{\dot{x}}(\omega)=(\omega_++\omega_-)d^2$,
 while the response spectrum of the velocity, measured via applying a small and periodic force, is given by
$ \tilde{R}_{\dot{x}}(\omega)=d^2[2\theta (w_+-w_-) + (w_++w_-)]/2T$. 
 These spectra are constant in the frequency domain due to the translational invariance of this simple model.   Indeed,  the FDT is violated even in the high frequency limit due to the presence of asymmetry,  as illustrated in Fig.~\ref{fig:motor2}(a).   
 
 The relative high-frequency violation  $\mathcal{V}_\infty/\tilde{C}_{\dot{x}}(\infty)$ becomes smaller when the  entropic change  per transition, i.e., $\Delta U/T$, decreases [Fig.~\ref{fig:motor2}(b)].    It can be checked easily that  Eq.~(\ref{eq:asymmetry-vio}) holds here, as illustrated in  Fig.~\ref{fig:motor2}(c).  Therefore,  a smaller driving energy $\Delta U$ reduces $\mathcal{V}_{\infty}$ by slowing down the biased motion [Fig.~\ref{fig:motor2}(b)].   Such a violation has been noticed  recently in a more realistic model of kinesin~\cite{ariga2017nonequilibrium}.   For a small $\Delta U/T$,  the renormalized FDT violation integral $I_*$ multiplied with the effective friction $\gamma_*$ well predicts the dissipation rate $\dot{q}$, as shown in Fig.~\ref{fig:motor2}(b).

 \emph{Practical renormalization.--} An important parameter in applying the HS equality is the ambient temperature $T$, which  is  very challenging to determine (or control) experimentally due to the tiny size of the molecular machine.  In practice,  $T$ has been determined from the ratio $\tilde{C}_{\dot{x}}(\omega)/2\tilde{R}_{\dot{x}}'(\omega)$ in the high-frequency regime, assuming that FDT is satisfied there~\cite{toyabe2010nonequilibrium,toyabe2015single,ariga2017nonequilibrium}.  According to our current study, this assumption might be wrong in the presence of perturbation asymmetry.   In fact,  the high-frequency violation leads to a modified temperature: 
    \begin{equation}
 T_{re}\equiv \lim_{\omega\to \infty} \frac{\tilde{C}_{\dot{x}}(\omega)}{2\tilde{R}_{\dot{x}}'(\omega)}=\frac{T}{1-\mathcal{V}_\infty/\tilde{C}_{\dot{x}}(\infty)}.
 \end{equation}
 It is reduced to the bath temperature when $\mathcal{V}_\infty=0$. 
 With this  temperature, we obtain a renormalized FDT violation integral that becomes  well-behaved:
 \begin{equation}
I_{re} \equiv  \langle \dot{x}\rangle_{ss}^2+  \int_{-\infty}^\infty\Big[ \tilde{C}_{\dot{x}}(\omega)-2T_{re} \tilde{R}_{\dot{x}}'(\omega)\Big]\frac{d\omega}{2\pi}.
\label{eq:I-original}
\end{equation}
  The effective friction coefficient $\gamma_{re}$ can be determined from 
 \begin{equation}
 \gamma_{re} \equiv \lim_{\omega\to \infty} \frac{1}{\tilde{R}_{\dot{x}}'(\omega)},
 \label{eq:gamma}
 \end{equation}
 which is an exact relation for Langevin systems, and serves as a generalization here.  When the entropic change per jump is small,  we have $\mathcal{V}_\infty/\tilde{C}_{\dot{x}}(\infty)\ll 1$, thus $I_{re}\approx I_{*}$  and $\gamma_{re}\approx \gamma_*$.   Therefore,   $\gamma_{re} I_{re}$ also becomes a reasonable estimation of the dissipation rate  $\dot{q}$ when the entropic change per jump is small. This is illustrated numerically in Supplemental Material using the 1-d hopping model~\cite{supp}.   When the high-frequency FDT violation is relatively  small,  different renormalization schemes converge to the same correct answer, although the original HS equality still breaks down due to the divergence of the FDT violation integral $I$.


\emph{Conclusion.--}We have demonstrated  for  Markov systems   that  the FDT violation persists generally in the high frequency limit in the presence of   asymmetric perturbation. This is in sharp contrast to our physical intuition that the high-frequency correlation and response essentially reflect only the thermal property of the bath.   
The high-frequency violation leads to a divergent FDT violation integral that invalidates the HS equality.  However, proper renormalization of the FDT violation integral restores the HS equality effectively when the entropic change in the environment  is small for each jump.  Hence, our study provides a protocol to estimate the dissipation rate for discrete Markov systems with asymmetry, based on the measured correlation and response spectra.   Our study also reveals a  linear relation between the high-frequency-limit violation and the asymmetric factor $\theta$, and therefore can be exploited to infer $\theta$ experimentally.  	We believe that our results will  guide further investigation of kinesin~\cite{ariga2017nonequilibrium}.

\begin{acknowledgements}
The author thanks Kyogo Kawaguchi  for motivating this project and helpful suggestions.  The author also thanks  Ben Machta for helpful suggestions.   The work was partially supported by the NSFC under Grant No. U1430237 and 11635002. 
\end{acknowledgements}


\begin{thebibliography}{10}
\bibitem{note} 
We note that $\langle \dot{Q}\rangle_{ss}=0$ in this system due to its  finite state space.  However, we expect  Eq.~(\ref{eq:lambda-beta-phi-3}) to work even for systems with an infinite state space,  as suggested by our analytical and numerical examples.
\end{thebibliography}

\onecolumngrid

\def\theequation{S\arabic{equation}}
\makeatletter
\@addtoreset{equation}{section}
\makeatother

\setcounter{equation}{0}

\onecolumngrid
\def\theequation{S\arabic{equation}}
\makeatletter
\@addtoreset{equation}{section}
\makeatother
\setcounter{equation}{0}

\def\thefigure{S\arabic{figure}}
\makeatletter
\@addtoreset{figure}{section}
\makeatother
\setcounter{figure}{0}
\newpage

\section*{Supplemental Material}


%

\subsection{Correlation, response, and FDT violation in general Markov models}

 \subsubsection{Correlation spectrum}
Noting that  the correlation function for $Q_t$ satisfies $ C_{Q}(t-\tau)\equiv \langle [Q_t-\langle Q\rangle_{ss}][Q_\tau-\langle Q\rangle_{ss}]\rangle_{ss}$,  we have 
\begin{equation}
 C_{\dot{Q}}(t-\tau)=\frac{\partial^2 C_{Q}(t-\tau)}{\partial \tau\partial t}.
 \label{eq:C-transform}
 \end{equation}
It is easier to calculate $C_{Q}(t-\tau)$ first.  Assuming $t\ge \tau$,  it satisfies
\begin{equation}
C_{Q}(t-\tau)= \sum_{n,n'}  \mathcal{Q}_{n}\mathcal{Q}_{n'} P(t-\tau;n,n')P_{n'}^{ss} -\langle Q\rangle_{ss}^2,
\label{eq:CQ}
\end{equation}
where $P(t-\tau;n,n')$ is the propagator,  or the probability for reaching state $n$ at time $t$,  assuming that the system starts from state   $n'$ at time $\tau$.  In the eigenspace,   
\begin{equation}
P(t-\tau;n,n')=\sum_{j} y_j(n') e^{-\lambda_j |t-\tau|} x_j(n).
\end{equation}
Indeed,  it is the solution of the corresponding master equation~(\ref{eq:govMatrix}) in the Main Text,  given the initial condition $P(0;n,n')=\delta_{nn'}$.   Inserting this relation back to  Eq.~(\ref{eq:CQ}) and introducing the projection of  $\mathcal{Q}$ on the $j$-th eigenmode,   i.e.,  $\alpha_j\equiv\sum_n \mathcal{Q}_nx_j(n)$ and $\beta_j \equiv \sum_{n}\mathcal{Q}_{n}  y_j(n)P_n^{ss}$,  we obtain   the expansion of correlation function   in the eigenspace: 
\begin{equation}
C_{Q}(t-\tau) =\sum_{j=2}^N \alpha_j \beta_j e^{-\lambda_j |t-\tau|}.
\label{eq:Corr1}
\end{equation}  
The contribution of the first eigenmode is counteracted by  $\langle Q\rangle^2_{ss}$.   Stationarity of the system guarantees that    $C_Q(t-\tau)=C_Q(\tau-t)$.  Therefore, Eq.~(\ref{eq:Corr1}) obtained from $t\ge \tau$ is also applicable for $t<\tau$.   We use the following convention for Fourier transform: 
\begin{equation}
\tilde{f}(\omega)=\int_{-\infty}^{\infty} f(t)\exp(i\omega t)dt,\quad 
f(t)=\int_{-\infty}^{\infty} \tilde{f}(\omega)\exp(-i\omega t) \frac{d\omega}{2\pi}.
\end{equation}   
Combining  Eq.~(\ref{eq:C-transform}), Fourier transformation and Eq.~(\ref{eq:Corr1}),   we finally obtain the velocity correlation spectrum~(\ref{eq:high-FDT-general}a) in the Main Text.

\subsubsection{Response spectrum}
The response spectrum can be obtained by studying the response of the system to   a  periodic perturbation.  Consider  $h_t=h_0 \exp(-i \omega t  )$ with $h_0$ a small amplitude and $ i$ the imaginary unit.  Expanded in Taylor series,  the modified  transition rate matrix $\tilde{M}$ is given by 
\begin{equation}
\tilde{M}=M+ M^* h_0 \exp(-i \omega t  )+O(h_0^2),
\end{equation}
where $M^*\equiv  \partial_h \tilde{M}|_{h\to 0}$.  On the other hand,    the modified distribution can also be expanded up to the first order:  
\begin{equation}
\tilde{P}_m=P_m^{ss}+ P^*_m h_0 \exp(-i \omega t  )+O(h_0^2),
\end{equation}
with $P_m^*\equiv  \partial_h \tilde{P}_m|_{h\to 0}$.   Since $d\tilde{P}_m/dt=\sum_n \tilde{M}_{mn}\tilde{P}_n$ and $\sum_n M_{mn}P_n^{ss}=0$,  we obtain in a Matrix form
 \begin{equation}
 P^{*}=-\frac{1}{M+i\omega  }M^* P^{ss}.
 \end{equation}
 For the observable $Q_t$,   its response spectrum is given by 
\begin{equation}
\tilde{R}_Q(\omega)=\sum_n \mathcal{Q}_nP^*_n=\sum_{j=2}^N \frac{\alpha_j\phi_j}{\lambda_j-i\omega},
\label{eq:Rq-general-fre}
\end{equation}
where $\phi_j \equiv \sum_{n} B_{n} y_j(n)$ with  $ B_n\equiv\sum_{m} M^*_{nm} P^{ss}_m$.   By using the transformation $ R_{\dot{Q}}(t)=dR_{Q}/dt$ or $\tilde{R}_{\dot{Q}}(\omega)=- i\omega \tilde{R}_{Q}(\omega)$,   we obtain the velocity response spectrum~(\ref{eq:high-FDT-general}b) in the Main Text.

\subsubsection{The renormalized FDT violation integral $I_*$}

 We derive Eq.~(\ref{eq:lambda-beta-phi-3}) in the Main Text.   Firstly, note that $(T\phi_j-\lambda_j \beta_j)$ is a key quantity in the violation spectrum integral:
 \[I_*=\sum_j \lambda_j\alpha_j \left(T\phi_j-\beta_j\lambda_j\right).\]
    According to definitions of these coefficients,  we obtain 
\[ T\phi_j-\lambda_j\beta_j =\sum_{n,m}y_j(n)J_m^n\Big(\frac{\mathcal{Q}_n+\mathcal{Q}_m}{2}+\theta_n^m(\mathcal{Q}_m-\mathcal{Q}_n)\Big). \]
For equilibrium systems,  the flux $J_m^n$ vanishes  due to detailed balance.   This leads to $T\phi_j=\beta_j\lambda_j$ for all eigenmodes,  and thus the  vanishing of the FDT violation integral.   On the other hand,   $\lambda_j\alpha_j =-\sum_n \bar{\nu}_n x_j(n)$, with  $\bar{\nu}_n\equiv \sum_m w_n^m (\mathcal{Q}_m-\mathcal{Q}_n)$  being the average change rate of $Q_t$  when it starts from state $n$.   Combining these relations,  we  obtain the analytical expression for the effective  FDT violation integral:
\begin{equation}
I_*=\sum_{n,m}\bar{\nu}_nJ_n^m\Big(\frac{\mathcal{Q}_n+\mathcal{Q}_m}{2}+\theta_n^m(\mathcal{Q}_m-\mathcal{Q}_n)\Big).
\label{eq:Delta-Q-1}
\end{equation}
Noting that $\sum_m J_n^m=0$ due to stationarity,  we  can subtract $\sum_{n,m}\bar{\nu}_nJ_n^m \mathcal{Q}_n$ (which is also zero) from  $I_*$, and  symmetrize the resulting expression to obtain Eq.~(\ref{eq:lambda-beta-phi-3}) in the Main Text.

\subsection{The renormalized HS equality}
Here, we provide more details of deriving the renormalized HS equality, give  numerical illustrations,  and present the generalization to higher dimensional models mentioned in the Main Text. First, we consider the 1-d hopping model mentioned in Fig.~\ref{fig:continuum}(a).  Following Eq.~(\ref{eq:I_*-1d}),  we are interested in how $\bar{\nu}_n+\bar{\nu}_{n+1}$ and $\bar{\nu}_n-\bar{\nu}_{n+1}$ behave when $\epsilon_n$ is small.  Here, $\bar{\nu}_n=d(w_n^+-w_n^-)$.  More explicitly, we have 
\begin{equation*}
\begin{split}
\bar{\nu}_n+\bar{\nu}_{n+1}&=w_0d \Big[(e^{(\theta+1/2)\epsilon_n}-e^{(\theta-1/2)\epsilon_{n-1}})+ (e^{(\theta+1/2)\epsilon_{n+1}}-e^{(\theta-1/2)\epsilon_{n}} )\Big].
\end{split}
\end{equation*}
Applying Taylor expansion, we obtain 
\begin{equation*}
\begin{split}
\bar{\nu}_n+\bar{\nu}_{n+1}&= w_0d \Big[ \epsilon_n+\frac{\epsilon_{n-1}+\epsilon_{n+1}}{2}+ \theta  (\epsilon_{n+1}-\epsilon_{n-1})\\
&+\theta (\epsilon_n^2+\frac{\epsilon_{n+1}^2+\epsilon_{n-1}^2}{2} )+(\frac{1}{8}+\frac{1}{2}\theta^2)(\epsilon_{n+1}^2-\epsilon_{n-1}^2) + O(\epsilon^3) \Big],
\end{split}
\end{equation*}
with $\epsilon$ capturing the overall amplitude of $\epsilon_n$. Similarly,  we have
\begin{equation*}
\begin{split}
\bar{\nu}_n-\bar{\nu}_{n+1}&=w_0d \Big[\frac{\epsilon_{n-1}-\epsilon_{n+1}}{2}+2\theta (\epsilon_n-\frac{\epsilon_{n+1}+\epsilon_{n-1}}{2})\\
&+ (\frac{1}{4}+\theta^2) (\epsilon_n^2-\frac{\epsilon_{n-1}^2+\epsilon_{n+1}^2}{2})-\frac{\theta}{2}(\epsilon_{n+1}^2-\epsilon_{n-1}^2)+ O(\epsilon^3)\Big].
\end{split}
\end{equation*}
Now, we assume $\epsilon_n=\epsilon f(nd)$, with $f(x)$ a continuous function and $\epsilon\propto d$.  The motivation of this assumption is that there is an underlying smooth energy landscape, as discussed in the Main Text.   From Taylor expansion,  we obtain
\[\epsilon_{n\pm1}=\epsilon \Big[ f(nd) \pm \frac{\partial f}{\partial x} d +\frac{1}{2}\frac{\partial^2 f}{\partial x^2}d^2+O(\frac{d^3}{L^3}) \Big]. \]
Therefore,  we have 
\begin{eqnarray}
\epsilon_{n+1}+\epsilon_{n-1}&=&2\epsilon_n+O\Big(\epsilon \frac{d^2}{L^2}\Big),\\
\epsilon_{n+1}-\epsilon_{n-1}&=& 2 \epsilon \frac{\partial f}{\partial x} d +O\Big(\epsilon \frac{d^3}{L^3}\Big),\\
\epsilon_{n+1}^2+\epsilon_{n-1}^2&=& 2\epsilon_n^2 + O\Big(\epsilon^2 \frac{d^2}{L^2}\Big),\\ 
\epsilon_{n+1}^2-\epsilon_{n-1}^2&=& 4 \epsilon \epsilon_n \frac{\partial f}{\partial x}d +O\Big(\epsilon^2 \frac{d^3}{L^3}\Big).
\end{eqnarray}
 Plugging  these relations back into the Taylor expansion of $\bar{\nu}_n+\bar{\nu}_{n+1}$ and $ \bar{\nu}_n-\bar{\nu}_{n+1}$, and noting that $d/L=O( \epsilon)$,  we finally obtain
 \begin{eqnarray}
\bar{\nu}_n+\bar{\nu}_{n+1} &=& 2w_0d \Big(\epsilon_n+\theta O(\epsilon^2)+O(\epsilon)^3\Big),\\
 \bar{\nu}_n-\bar{\nu}_{n+1}&=& O(\epsilon^2) .
 \end{eqnarray}
 This then leads to Eq.~(\ref{eq:GHS}) in the Main Text.

\begin{figure}
 \centering
 \includegraphics[width=14cm]{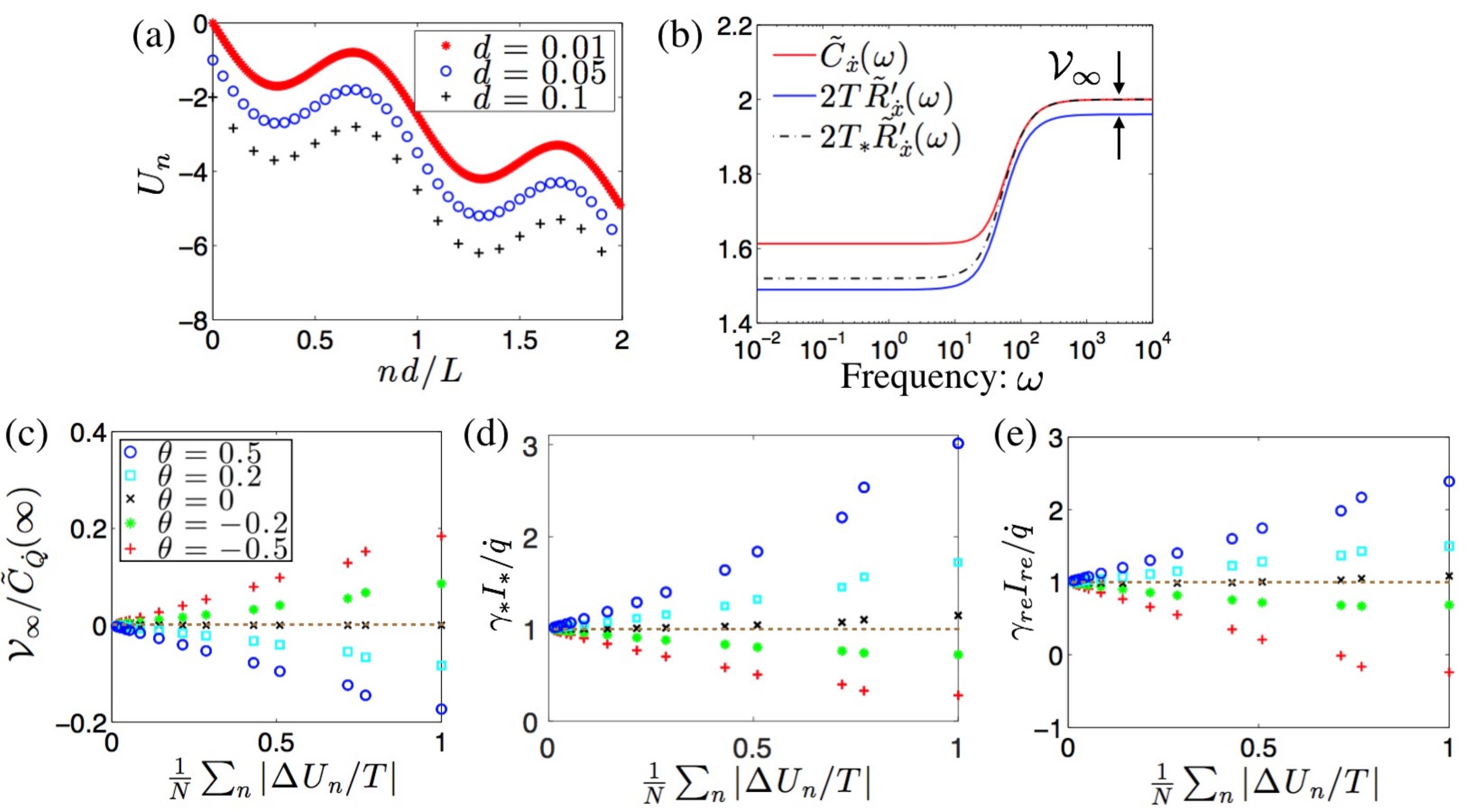}
 \caption{ (a) The  energy landscape $U_n$ for the 1-d hopping model,  constructed from $U(x)=-\sin (2\pi x/L)-\Delta \mu x/L$ at a given lattice constant $d$.  The landscapes  for different $d$'s are shifted vertically for illustration.  (b) The correlation and response spectrum  for the velocity $\dot{x}_{t}\equiv \dot{n}_t d$, obtained at $d=0.1$ and $\theta=-0.2$.  The FDT is restored in the high frequency limit with the renormalized temperature.  (c)  The relative high-frequency-limit violation of FDT against the average entropic change in the environment per jump: $\frac{1}{N}\sum_n |\Delta U_n/T|$.    (d)(e)  Emergence of the  renormalized HS equality at small medium entropy production per jump.     Other parameters: $w_0=1/d^2$, $T=1$, $L=1$, $\Delta \mu=2.5$, and $h=0$.  }
 \label{fig:continuum-S}
 \end{figure}

We provide a numerical illustration  for Eq.~(\ref{eq:GHS}). Consider that $U(x)=U(x+L)+\Delta \mu$ is a periodic function tilted by an energy input $\Delta \mu$ in each period $L$,  which drives the system out of equilibrium.  This is illustrated  in    Fig.~\ref{fig:continuum-S}(a).   The number of states within each period is $N=L/d$.   The prefactor $w_0$ scales with $1/d^2$ so that the global features (mean velocity etc) converge to a finite value in the continuum limit $d/L\to 0$.     The correlation and response spectrum is shown in FIG.~\ref{fig:continuum}(b) for $d=0.1$ and $\theta=-0.2$.   Again, the FDT violation persists even in the high frequency limit.   The average entropy production in the environment per jump,  $\frac{1}{N}\sum_n|\Delta U_n/T|$, is a crucial parameter here.   It roughly scales with the discreteness $d$ of the system, and vanishes in the limit $d/L\to 0$.  By changing the discreteness in our numerical simulation,  we find that,  below a sufficiently small $\frac{1}{N}\sum_n|\Delta U_n/T|$,   the relative high-frequency FDT violation becomes negligible [Fig.~\ref{fig:continuum-S}(c)], and the renormalized HS equality  emerges  [Fig.~\ref{fig:continuum-S}(d)].  This holds true for various values of $\theta$.  Again, $\theta=0$ is special in that the corresponding $\mathcal{V}_\infty=0$, and $\gamma_{*} I_{*}$ proves to be a much more accurate (though not exact) estimation for the dissipation rate $\dot{q}$ [Fig.~\ref{fig:continuum-S}(d)].  Another renormalization scheme based on a modified temperature has similar properties [Fig.~\ref{fig:continuum-S}(e)].

Finally, we focus on the multi-dimensional hopping models in Fig.~\ref{fig:continuum}(b)(c) in the main text,  where the same value of $\mathcal{Q}_p=pd$ is shared by all the states within the same colored block.    The perturbed rates of the  \emph{red} transitions that change the observable are assumed to satisfy 
\begin{subequations}\label{eq:interblock-rates}
 \begin{eqnarray}
\tilde{w}_m^{n}&=&w_0\exp\Big((\theta+\frac{1}{2} ) \Big[\epsilon_m^n+h\frac{\mathcal{Q}_n-\mathcal{Q}_m}{T}\Big] \Big),\\
\tilde{w}_n^{m}&=&w_0\exp\Big((\theta-\frac{1}{2} ) \Big[\epsilon_m^n+h\frac{\mathcal{Q}_n-\mathcal{Q}_m}{T}\Big] \Big),
\end{eqnarray}
\end{subequations}
which essentially mimics Eq.~(\ref{eq:w-theta-1d}), except that we do not assume an energy landscape $U_n$.   The dissipation rate through the stochastic trajectory $Q_t$ is defined to be 
\begin{equation}
\dot{q}\equiv T \sum_{n,m}(1-\delta_{\mathcal{Q}_n\mathcal{Q}_m})J_n^m \epsilon_n^m, 
\label{eq:diss-Q}
\end{equation}
where $\epsilon_n^m=\ln [\omega_n^m/\omega_m^n]$ is the  environment's entropy production for the transition from state $n$ to $m$,  and   $(1-\delta_{\mathcal{Q}_n\mathcal{Q}_m})$ is a weight that only counts   transitions that change the observable.   Assuming that both   $\epsilon_n^m$ and  its relative variation  are small along the direction of red transitions,  the renormalized HS equality also emerges.  The differences of  network topologies are captured by the effective friction coefficient $\gamma_*=4T/(kw_0d^2)$,  with $k$ being the number of red transitions out of a node.

\end{document}